УДК 669.86:536.21

# ДЛИННОВОЛНОВАЯ НЕУСТОЙЧИВОСТЬ АДВЕКТИВНОГО ТЕЧЕНИЯ В НАКЛОННОМ СЛОЕ С ИДЕАЛЬНО ТЕПЛОПРОВОДНЫМИ ГРАНИЦАМИ

Сагитов Р.В., Шарифулин А.Н.

Пермский государственный технический университет,614990, Пермь

E-mail: sharifulin@pstu.ru

Исследуется устойчивость стационарного конвективного течения в плоском наклонном слое с идеально теплопроводными твердыми границами при наличии однородного продольного градиента температуры. Аналитически найдены границы устойчивости по отношению к длинноволновым возмущениям, найдены критические числа Грасгофа для наиболее опасных среди них четных спиральных возмущений.

*Ключевые слова*: Адвективное течение, наклонный слой, продольный градиент температуры, длинноволновая неустойчивость

## 1. Введение

Интерес к течениям в бесконечных слоях, обусловленным продольным градиентом температуры, связан с рядом геофизических и технологических приложений. К ним относятся, например, горизонтальные адвективные течения в атмосфере и океане, конвекция в вертикальных и наклонных шахтных выработках и нефтяных скважинах. Такого рода исследования начаты Остроумовым Г.А. [1], сформулировавшим задачу об устойчивости равновесия бесконечного слоя жидкости между твердыми вертикальными идеально теплопроводящими пластинами, на которых поддерживается направленный вниз постоянный градиент температуры. Результаты исследования его устойчивости для различных вариантов граничных условий приведены в монографии [2]. В [1] эта задача была обобщена на случай наклона слоя и наличия кроме продольного также и поперечного градиента температуры. Там же получены точные аналитические выражения для профилей скорости и температуры для случая, когда ось наклона горизонтальна, а продольный и поперечный градиенты температуры перпендикулярны к ней. Многие предельные случаи этой задачи (подогреваемый снизу плоский слой, вертикальный слой, подогреваемый сбоку при наличии или отсутствии продольного градиента, наклонный слой между изотермическими пластинами) в дальнейшем были детально исследованы на устойчивость [2,3].

Для случая, когда слой горизонтален и отсутствует поперечный градиент температуры в [4] получены простые выражения для профилей скорости и температуры плоскопараллельного течения для четырех различных вариантов граничных условий. Результаты исследования устойчивости этих и течений, различными авторами приведены в [3]. Подробный анализ устойчивости конвективного течения для случая, когда обе стенки слоя твердые приведен в [5]. Обзор работ, связанных с новыми постановками задач, где основное течение порождается продольным градиентом температуры, можно найти в [6].

Исследование устойчивости плоскопараллельного конвективного течения в наклоненном слое начато R. Delgado-Buscalioni и E. Crespo del Arco [7] . В их работе рассмотрена устойчивость течения в наклонном слое с продольным градиентом для случая теплоизолированных границ. В настоящей работе приводятся результаты аналитического исследования устойчивости конвективного течения по отношению в длинноволновым возмущениям в наклоненном плоском слое с продольным градиентом температуры для случая теплопроводных границ, которое ранее не проводилось.

## 2. Постановка задачи.

Рассмотрим наклоненный плоский бесконечный слой жидкости толщиной $2d$, ограниченный идеально теплопроводными твердыми параллельными плоскостями, на которых задан постоянный градиент температуры $\vec{A}$ (см. Рис. 1). Введем декартову систему координат, жестко связанную со слоем: начало системы координат расположено в центре слоя, ось $x$ направлена перпендикулярно стенкам, ось $y$ – горизонтальна, а ось $z$ совпадает с направлением $\vec{A}$. Слой наклонен так, что ускорение свободного падения $\vec{g}$ лежит в плоскости $(x,z)$ и образует с осью $z$ угол $\alpha$.

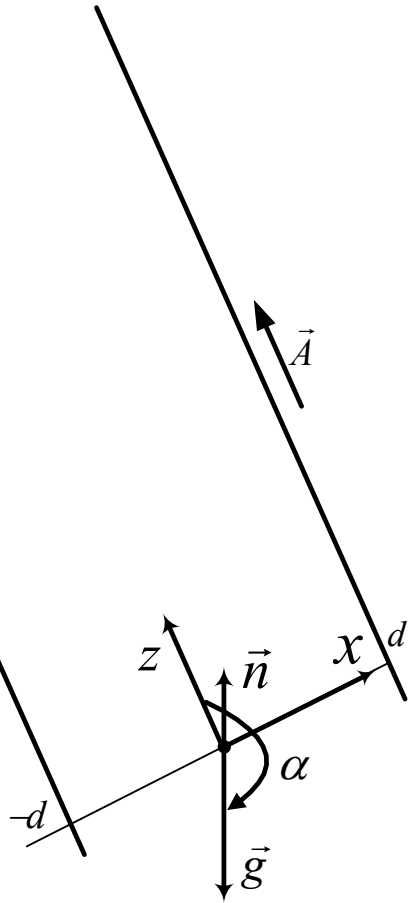

**Рис. 1** Наклоненный слой с условиями подогрева и осями координат.

Границы полости твердые и идеально теплопроводны. На них поддерживается постоянный градиент температуры $\vec{A}$, так что температура линейно растет вдоль оси $z$

$$T^{(0)} = z \tag{1}$$

Для описания движения жидкости используем уравнения свободной тепловой конвекции в приближении Буссинеска, которые в безразмерной форме запишутся в виде:

$$\frac{\partial \vec{v}}{\partial t} - \vec{v} \times (\nabla \times \vec{v}) = -\nabla(p + \frac{\vec{v}^2}{2} + Gr \cdot zx \sin\alpha + Gr \cdot \cos\alpha \frac{z^2}{2}) - \nabla \times (\nabla \times \vec{v}) + \\ + Gr\vec{n}T + Gr \cdot x\sin\alpha \cdot \vec{e}_z, \tag{2}$$

$$\frac{\partial T}{\partial t} + \vec{v} \cdot \nabla T + v_z = \frac{1}{Pr}\Delta T, \tag{3}$$

$$\nabla \cdot \vec{v} = 0, \tag{4}$$

В систему уравнений (2) –(4) входят 3 безразмерных параметра:

- число Грасгофа $Gr = \dfrac{A\beta g d^4}{\nu^2}$,

- число Прандтля $Pr = \dfrac{\nu}{\chi}$,

- угол наклона слоя $\alpha$.

В дальнейшем будем использовать и число Рэлея $Ra = Gr \cdot Pr$.

Граничные условия запишем в виде:

$$x = \pm 1: \vec{v} = 0, T = 0. \tag{5}$$

Здесь $\vec{v}$ – скорость, $T$ –отклонение температуры от распределения (1), реализующегося в теплопроводном режиме в отсутствие движения жидкости, $p$ – давление. $\vec{g} = -\vec{n}g = (g\sin\alpha, 0, g\cos\alpha)$ – вектор ускорения свободного падения, $\vec{n} = (-\sin\alpha, 0, -\cos\alpha)$ - единичный вектор, направленный вверх, $\vec{e}_z$ - орт оси $z$.

Считаем, что полость замкнута, поэтому расход жидкости через любое сечение равен нулю. Выбираем сечение перпендикулярно оси $z$, тогда условие замкнутости приобретет вид

$$\int_{-\infty}^{+\infty}\int_{-1}^{1} v_z \, dx \, dy = 0. \tag{6}$$

В качестве единиц измерения расстояния, времени, скорости, температуры и давления выбраны, соответственно $d$, $d^2/\nu$, $\nu/d$, $Ad$ и $\rho\nu^2/d^2$; $\rho$ – средняя плотность жидкости, $\beta$ – объемный коэффициент теплового расширения, $\nu$ – кинематическая вязкость, $\chi$ – коэффициент температуропроводности.

### *3. Плоскопараллельное течение*

Задача (2) – (6) имеет решение $\vec{v}=(0,0,\mathrm{v}_0)$, $T=\vartheta_0$, соответствующее стационарному плоскопараллельному течению с единственной $z-$составляющей скорости $\mathrm{v}_0$:

$$\mathrm{v}_0 = \upsilon_0 Gr\sin\alpha, \quad \vartheta_0 = \tau_0 \cdot \mathrm{tg}\,\alpha, \qquad (7)$$

где при $|\alpha| \leq 90^0$

$$\upsilon_0 = \frac{1}{2\gamma^2}\left(\frac{\sin\gamma x}{\sin\gamma} - \frac{\sinh\gamma x}{\sinh\gamma}\right), \qquad (8)$$

$$\tau_0 = x - \frac{1}{2}\frac{\sin\gamma x}{\sin\gamma} - \frac{1}{2}\frac{\sinh\gamma x}{\sinh\gamma}, \qquad (9)$$

$$\gamma = \sqrt[4]{PrGr\cos\alpha}, \qquad (10)$$

при $|\alpha| \geq 90^0$:

$$\upsilon_0 = \frac{\cos\zeta\sinh\zeta\sin\zeta x\cosh\zeta x - \sin\zeta\cosh\zeta\cos\zeta x\sinh\zeta x}{2\zeta^2\left(\cos^2\zeta - \cosh^2\zeta\right)}, \qquad (11)$$

$$\tau_0 = x + \frac{\cosh\zeta\sin\zeta\cosh\zeta x\sin\zeta x + \sinh\zeta\cos\zeta\sinh\zeta x\cos\zeta x}{\cos^2\zeta - \cosh^2\zeta}, \qquad (12)$$

$$\zeta = \sqrt[4]{-\frac{1}{4}\Pr Gr\cos\alpha}. \tag{13}$$

В случае $\alpha = 90°$ (слой горизонтален) решения (7),(8),(9) и (7),(11),(12) могут быть записаны в виде:

$$v_0 = \left(x - x^3\right)\frac{Gr}{6}, \quad \vartheta_0 = \left(-\frac{7}{30}x + \frac{1}{3}x^3 - \frac{1}{10}x^5\right)\frac{GrPr}{12}. \tag{14}$$

Решение (14) получено в [2] и часто [8,6] называется течением Бириха, а соответствующая задача, – задачей Бириха. Его устойчивость хорошо исследована, как теоретически [3, 5], так и экспериментально [12]. В связи с технологическими приложениями, активно исследуется влияние на его структуру и устойчивость вращения[11], вибраций полости[9,10], бинарных свойств жидкости[13,14].

Отклонение положения слоя от горизонтального приводит к качественному изменению вида плоскопараллельного течения.

**Подогрев снизу.** В случае $0 < |\alpha| < 90°$ с увеличением числа Грасгофа $Gr$ при фиксированных значениях $\alpha$ и $Pr$ при переходе через значение $Gr = Gr_*^{(n)}$, соответствующих $\gamma = \pi n$ происходит смена знака скорости $v_0$ и температуры $\vartheta_0$ плоскопараллельного течения. Эти инверсионные значения числа Грасгофа определяются соотношением:

$$Gr_{inv}^{(n)} = \frac{(n\pi)^4}{Pr\cos\alpha}, \tag{15}$$

где $n$ - натуральное число.

Профили скорости $v_0$ и температуры $\vartheta_0$ для некоторых значений $\gamma$ представлены представлены на Рис. 2 и Рис. 3.

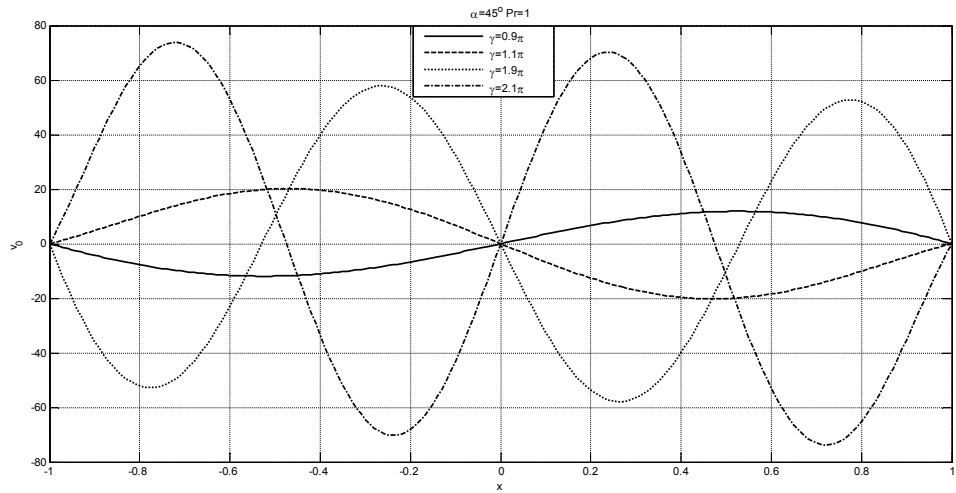

**Рис. 2** Профиль скорости $v_0$ для некоторых значений параметра $\gamma$ при $\alpha = 45°, Pr = 1$.

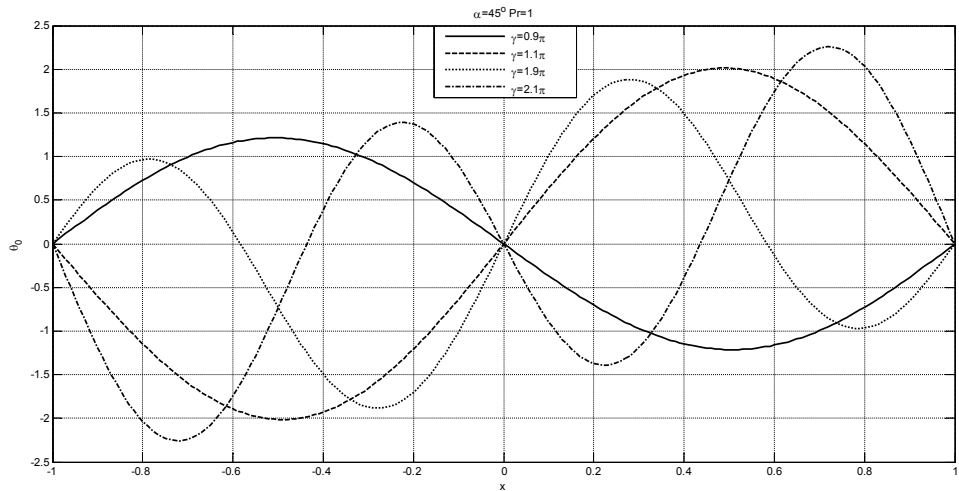

**Рис. 3** Профили температуры $\theta_0$ для некоторых значений параметра $\gamma$ при $\alpha = 45°, Pr = 1$.

Видно, что при увеличении $Gr$ в распределении скорости $v_0$ и температуры $\vartheta_0$ появляются новые узлы. При $Gr = Gr_{inv}^{(n)}$ скорость и температура плоскопараллельного течения обращаются в бесконечность.

Бесконечная интенсивность течения (7)-(9) не может быть реализована хотя бы потому, что в этом случае через единицу объема жидкости будет проходить бесконечный поток тепла, для поддержания которого нужен источник энергии бесконечной мощности. Кроме того, как будет показано ниже, уже при умеренных значениях числа Грасгофа $Gr < Gr_{inv}^{(1)}$, плоскопараллельное течения становится не устойчивым.

**Подогрев сверху.** Профили $v_0$ и $\theta_0$ при $|\alpha| > 90^0$ для некоторых значений $\zeta$ представлены на Рис. 2 и Рис. 3.

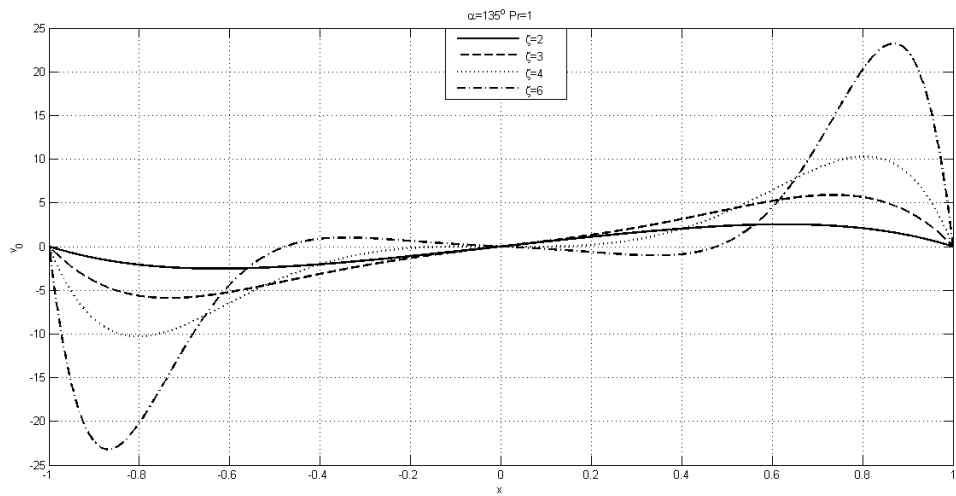

**Рис. 4** Профиль $v_0$ для некоторых значений параметра $\zeta$ при $\alpha = 135°, Pr = 1$.

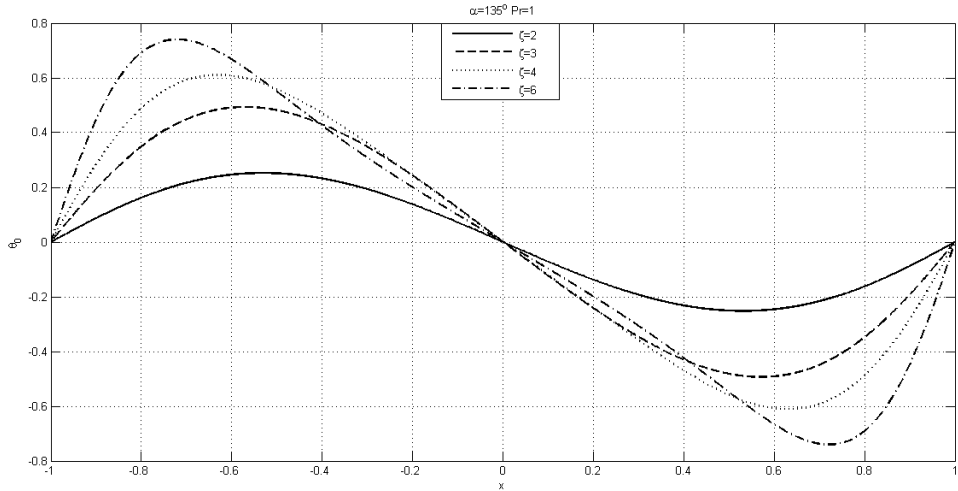

**Рис. 5** Профиль $\theta_0$ для некоторых значений параметра $\zeta$ при $\alpha = 135°, Pr = 1$.

При малых $\zeta$ профиль скорости близок к кубическому. С увеличением $\zeta$ течение в центральной области слоя замедляется. При $\zeta = \zeta_n$, приближённо определяемых формулой $\zeta_n \approx \frac{\pi}{4} + \pi n - \exp\left(-2\left(\frac{\pi}{4} + \pi n\right)\right)$, где $n$ – натуральное число, в распределении скорости в середине слоя появляются новые узлы. С увеличением $\zeta$ узлы смещаются от середины ближе к границе слоя. При этом течение жидкости между узлами довольно слабое, по сравнения с течением между границей и ближайшим к границе узлом. При $\zeta > 5$ жидкость течёт преимущественно в пограничных слоях толщиной порядка $\frac{\pi h}{\zeta}$.

**Малые возмущения.** Для исследования устойчивости течений (7)-(13) рассмотрим нестационарное возмущённое течение $\vec{v}_0 + \vec{v}(x,y,z,t)$,

$\theta_0 + \vartheta(x,y,z,t)$, $p_0 + p(x,y,z,t)$. Подставляя эти поля в (2)-(4) и линеаризуя по малым нестационарным возмущениям $\vec{v}(x,y,z,t)$, $\vartheta(x,y,z,t)$, $p(x,y,z,t)$ получаем систему уравнений для возмущений:

$$\frac{\partial \vec{v}}{\partial t} - \vec{v}_0 \times (\nabla \times \vec{v}) - \vec{v} \times (\nabla \times \vec{v}_0) = -\nabla(p + \vec{v}_0 \vec{v}) - \nabla \times (\nabla \times \vec{v}) + Gr\vec{n}\vartheta, \qquad (16)$$

$$\frac{\partial \vartheta}{\partial t} + \vec{v}_0 \cdot \nabla \vartheta + \vec{v} \cdot \nabla \theta_0 + v_z = \frac{1}{Pr}\Delta\vartheta, \qquad (17)$$

$$\nabla \cdot \vec{v} = 0, \qquad (18)$$

с граничными условиями

$$x = \pm 1 : \vec{v} = 0, \vartheta = 0. \qquad (19)$$

Рассмотрим два типа малых возмущений – плоские возмущения в виде валов с осями, параллельными оси $y$, и пространственные спиральные возмущения в виде валов с осями параллельными направлению основного потока.

## 4. *Длинноволновые возмущения*

Задача (16)-(19) имеет решения в виде плоских нормальных возмущений:

$$v_x(x,z,t) = \bar{v}_x(x)\exp(-\lambda t + ikz), \quad v_y = 0, \quad v_z(x,z,t) = \bar{v}_z(x)\exp(-\lambda t + ikz),$$
$$p(x,z,t) = \bar{p}(x)\exp(-\lambda t + ikz), \quad \vartheta(x,z,t) = \bar{\vartheta}(x)\exp(-\lambda t + ikz).$$

Эта же задача имеет решение и в виде спиральных возмущений:

$$v_x(x,y,t) = \tilde{v}_x(x)\exp(-\lambda t + iky), \quad v_y(x,y,t) = \tilde{v}_y(x)\exp(-\lambda t + iky),$$
$$v_z(x,y,t) = \tilde{v}_z(x)\exp(-\lambda t + iky),$$
$$p(x,y,t) = \tilde{p}_x(x)\exp(-\lambda t + iky), \quad \vartheta(x,y,t) = \tilde{\vartheta}_x(x)\exp(-\lambda t + iky).$$

Далее знаки тильда и верхнее подчеркивание у амплитуд возмущений будем опускать.

В длинноволновом пределе, т.е. при $k=0$, каждая задача имеет два типа решений: чётные и нечётные по $x$ относительно середины слоя. В случае чётного решения будем полагать амплитуды всех компонент скорости и температуры пропорциональными линейной комбинации $\cos(\sigma x)$ и $\mathrm{ch}(\sigma x)$, в случае нечётного – $\sin(\sigma x)$ и $\mathrm{sh}(\sigma x)$. Подставив эти соотношения в граничные условия (19), получаем:

$$\sigma = \pi \frac{m}{2}, \qquad (20)$$

Значения m=1,3,5…(m=2,4,6…) соответствуют чётным (нечётным) возмущениям.

Решение задачи при $k=0$ как для плоских нормальных, так и для спиральных возмущений дает выражение для декремента длинноволновых возмущений:

$$\lambda = \frac{\sigma^2(1+Pr) \pm \sqrt{\sigma^4(1-Pr)^2 + 4Pr^2 Gr\cos\alpha}}{2Pr}. \qquad (21)$$

Рассмотрим отдельно случаи отсутствия (присутствия) подогрева снизу.

**Подогрев снизу отсутствует** $|\alpha| \geq 90^0$. В этом случае, как это следует из (21), возмущения затухают. При значениях числа

Грасгофа $Gr > Gr_{oscil}$ возможны затухающие колебательные возмущения. Значение $Gr_{oscil}$ определяется соотношением:

$$Gr_{oscil} = -\sigma^4 \frac{(1-Pr)^2}{4Pr^2 \cos\alpha}. \qquad (22)$$

Декремент их затухания $\lambda_r$ и частота $\omega = \text{Im}(\lambda)$ определяются соотношениями:

$$\lambda_r = \sigma^2 \frac{1+Pr}{2Pr}, \quad \omega^2 = (Gr_{oscil} - Gr)\cos\alpha..$$

В соответствии с (22) при приближении угла $\alpha$ наклона к значению $90^0$ со стороны $\alpha = 180^0$ линия возникновения колебательных возмущений при $Pr \ne 1$ стремится к бесконечности. Это согласуется с результатами [15], где в маломодовом приближении получено, что на аналогичной линии возникновения колебательных возмущений для замкнутой полости при $\alpha \approx 90^0$ имеется точка возврата.

Дальнейшее изложение соответствует подогреву снизу $|\alpha| < 90^0$. Из (21) следует, что длинноволновые колебательные возмущения при подогреве снизу невозможны.

Представим амплитуды длинноволновых монотонных возмущений в виде разложения по малому параметру $k$:

$$\vec{v} = \vec{v}^{(0)} + \vec{v}^{(1)} k + \ldots,\ p = p^{(0)} + p^{(1)} k + \ldots,\ \vartheta = \vartheta^{(0)} + \vartheta^{(1)} k + \ldots.$$

**Чётные плоские возмущения.** В случае чётных плоских возмущений в нулевом порядке $\vartheta^{(0)} = 0, \vec{v}^{(0)} = 0, \dfrac{dp^{(0)}}{dx} = 0$. В первом порядке ненулевые амплитуды нейтральных возмущений скорости и температуры определяются выражениями:

$$\mathrm{v}_z^{(1)} = \frac{i\gamma^2 p^{(0)}}{2PrGr\cos\alpha}\left(-\frac{\cos(\gamma x)}{\cos(\gamma)} + \frac{\cosh(\gamma x)}{\cosh(\gamma)}\right),$$

$$\vartheta^{(1)} = \frac{ip^{(0)}}{Gr\cos\alpha}\left(-1 + \frac{1}{2}\left(\frac{\cos(\gamma x)}{\cos(\gamma)} + \frac{\cosh(\gamma x)}{\cosh(\gamma)}\right)\right),$$

где $\gamma^4 = PrGr\cos\alpha$.

Нейтральные кривые определяются корнями трансцендентного уравнения

$$\tan\gamma = \tanh\gamma. \qquad (23)$$

Первые 3 решения которого –

$$\gamma_1 \approx 3.9266, \gamma_2 \approx 7.0686, \gamma_3 \approx 10.2102, \ldots$$

Таким образом, чётные плоские длинноволновые возмущения затухают пока

$$Ra < Ra_{plain}^{even}(1) = \frac{\gamma_1^4}{\cos\alpha} \approx \frac{237.7}{\cos\alpha}. \qquad (24)$$

Уравнение (23) совпадает с уравнением, полученным в [2] при исследовании устойчивости равновесия в вертикальном слое, т.е. при $\alpha = 0$, соответственно и критическое значение числа Рэлея (24) при $\alpha \to 0$ переходит в полученное в [2] критическое значение.

**Нечётные плоские и спиральные возмущения.** Нейтральные кривые, соответствующее этим типам возмущений, получаются из (21),(20) при m=2n:

$$Ra_{plain}^{odd}(n) = Ra_{spiral}^{odd}(n) = \frac{(n\pi)^4}{\cos\alpha}, \quad n=1,2,.... \qquad (25)$$

Это выражение в предельном случае $\alpha \to 0$ переходит в полученное для случая вертикального слоя в [2]. Нейтральная кривая наиболее опасных возмущений соответствует $n=1$:

$$Ra_{plain}^{odd}(1) = Ra_{spiral}^{odd}(1) = \frac{\pi^4}{\cos\alpha} \approx \frac{97.41}{\cos\alpha}. \qquad (26)$$

Рассмотрев разложения следующего порядка малости по $k$ можно показать, что длинноволновые нечетные спиральные возмущения не являются наиболее опасными в сравнении с аналогичными имеющими малое, но конечное значение волнового числа.

**Чётные спиральные возмущения.** Уравнение семейства нейтральных кривых вытекает из (20), (21) при m=2n+1:

$$Ra_{spiral}^{even}(n) = \frac{\left(\pi\left(n+\frac{1}{2}\right)\right)^4}{\cos\alpha}, \quad n=0,1,2,.... \qquad (27)$$

Нейтральная кривая наиболее опасных возмущений соответствует $n=0$:

$$Ra_{spiral}^{even}(0) = \frac{\pi^4}{16\cos\alpha} \approx \frac{6.088}{\cos\alpha}. \qquad (28)$$

Рассмотрев разложения следующего порядка малости можно показать, что длинноволновые четные спиральные возмущения являются наиболее опасными в сравнении с аналогичными имеющими малое, но конечное значение волнового числа $k$.

## 5. Заключение

Аналитически изучена длинноволновая неустойчивость плоскопараллельного течения в наклонном слое с постоянным продольным градиентом температуры на его идеально теплопроводных границах. Получено, что в условиях подогрева сверху, т.е. при $|\alpha| > 90^0$, течение устойчиво по отношению ко всем типам длинноволновых возмущений, но при превышении числом Грасгофа значения, определяемого соотношением (22), возмущения затухают колебательным образом.

При $|\alpha| < 90^0$ плоскопараллельное течение теряет устойчивость при превышении числом Рэлея критического значения, зависящего от типа возмущения. Наиболее опасными среди длинноволновых возмущений являются четные спиральные возмущения, критическое значение числа Рэлея которых, определяется выражением (28). Показано, что они наиболее опасны и среди четных спиральных возмущений с малыми, но конечными значениями волнового числа.